\documentclass[twocolumn,amsmath,amssymb,superscriptaddress]{revtex4}
\usepackage{graphicx}
\usepackage{dcolumn}
\usepackage{bm}
\usepackage{color}

\begin{document}
\title{Highly excited exciton-polariton condensates}
\author{Tomoyuki Horikiri}
\thanks{\texttt{horikiri-tomoyuki-bh@ynu.ac.jp}}
\affiliation{Yokohama National University, Tokiwadai 79-5, Hodogaya-ku, Yokohaa, Kanagawa 240-8501, Japan} 
\affiliation{National Institute of Informatics, Hitotsubashi 2-1-2, Chiyoda-ku, Tokyo 101-8430, Japan}
\affiliation{E. L. Ginzton Laboratory, Stanford university, Stanford, 
California 94305, USA}
\affiliation{The University of Tokyo, 7-3-1 Hongo, Bunkyo-ku, Tokyo 113-8656, Japan}
\affiliation{Center for Emergent Matter Science, RIKEN, 2-1 Hirosawa, Wakoshi, Saitama 351-0198, Japan}
\author{Tim Byrnes}
\affiliation{State Key Laboratory of Precision Spectroscopy, School of Physical and Material Sciences,
East China Normal University, Shanghai 200062, China}
\affiliation{NYU-ECNU Institute of Physics at NYU Shanghai, 3663 Zhongshan Road North, Shanghai 200062, China}
\affiliation{National Institute of Informatics, 2-1-2 Hitotsubashi, Chiyoda-ku, Tokyo 101-8430, Japan}
\affiliation{New York University Shanghai, 1555 Century Ave, Pudong, Shanghai 200122, China}
\affiliation{Department of Physics, New York University, New York, NY 10003, USA}
\author{Kenichiro Kusudo}
\affiliation{National Institute of Informatics, Hitotsubashi 2-1-2, Chiyoda-ku, Tokyo 
101-8430, Japan}
\author{Natsuko Ishida}
\affiliation{Center for Emergent Matter Science, RIKEN, 2-1 Hirosawa, Wakoshi, Saitama 351-0198, Japan}
\affiliation{The 
University of Tokyo, 7-3-1 Hongo, Bunkyo-ku, Tokyo 113-8656, Japan}
\author{Yasuhiro Matsuo}
\affiliation{The University of Tokyo, 7-3-1 Hongo, Bunkyo-ku, Tokyo 113-8656, Japan}
\author{Yutaka Shikano}
\affiliation{Research Center of Integrative Molecular Systems, Institute for Molecular Science, 38 Nishigo-Naka, Myodaiji, Okazaki 444-8585, Japan}
\affiliation{Institute for Quantum Studies, Chapman University, 1 University Dr., Orange, California 92866, USA.}
\affiliation{Materials and Structures Laboratory, Tokyo Institute of Technology, 4259 Nagatsuta, Midori, Yokohama 226-8503, Japan.}
\affiliation{Research Center for Advanced Science and Technology (RCAST), The University of Tokyo, 4-6-1 Komaba, Meguro-ku, Tokyo, 153-8904, Japan}
\author{Andreas L$\ddot{\rm o}$ffler}
\affiliation{Technische Physik, Universit$\ddot{a}$t W$\ddot{u}$rzburg, Am 
Hubland, D-97074 W$\ddot{u}$rzburg, Germany}
\author{Sven H$\ddot{\rm o}$fling}
\affiliation{Technische Physik, Universit$\ddot{a}$t W$\ddot{u}$rzburg, Am 
Hubland, D-97074 W$\ddot{u}$rzburg, Germany}
\affiliation{SUPA, Schoold of Physics and Astronomy, University of St Andrews, KY16 9SS, United Kingdom.}
\affiliation{National Institute of Informatics, Hitotsubashi 2-1-2, Chiyoda-ku, Tokyo 
101-8430, Japan}
\author{Alfred Forchel}
\affiliation{Technische Physik, Universit$\ddot{a}$t W$\ddot{u}$rzburg, Am 
Hubland, D-97074 W$\ddot{u}$rzburg, Germany}
\author{Yoshihisa Yamamoto}
\affiliation{National Institute of Informatics, Hitotsubashi 2-1-2, Chiyoda-ku, Tokyo 101-8430, Japan}
\affiliation{E. L. Ginzton Laboratory, Stanford university, Stanford, 
California 94305, USA}
\affiliation{ImPACT Program, Japan Science and Technology Agency, 7 Gobancho, Chiyoda-ku, Tokyo 102-0076, Japan.}
\begin{abstract}
{Exciton-polaritons are a coherent electron-hole-photon (e-h-p) system where condensation has been observed in semiconductor microcavities. In contrast to
equilibrium Bose-Einstein condensation (BEC) for long lifetime systems, polariton condensates have a dynamical nonequilibrium feature owing to the similar physical structure that they have to semiconductor lasers. One of the distinguishing features of a condensate to a laser is the presence of strong coupling between the matter and photon fields. Irrespective of its
equilibrium or nonequilibrium nature, exciton-polariton have been observed to maintain strong coupling.  We show that by investigating high density regime of exciton-polariton condensates,
the negative branch directly observed in photoluminescence.  This is evidence that the
present e-h-p system is still in the strong coupling regime, contrary to past results where the system reduced to standard lasing at high density.
} 
\end{abstract}

\maketitle
An exciton-polariton is a half-matter, half-light quasiparticle that is formed by the strong coupling between quantum well excitons and microcavity photons. These quasiparticles undergo nonequilibrium, dynamical condensation \cite{snoke, huireview}, where a macroscopic population in the ground state and a long range phase order are spontaneously formed above the threshold pump rate. Despite the nonequilibrium and open-dissipative nature of the condensates, they exhibit many features in common with a thermal equilibrium Bose--Einstein condensate (BEC), such as superfluidity \cite{amo}, a quantized vortex \cite{lagoudakis}, the formation of vortex-antivortex pairs \cite{george}. 
In addition, excitation spectrum on both sides of the condensate energy \cite{utsunomiya,kohnle} has been observed by photoluminescence (PL) for the positive energy side \cite{utsunomiya} and for the negative energy side by pump-probe signal \cite{kohnle}, and direct PL \cite{pieczarka}.

 One of the attractive features of exciton-polaritons is that the condensate can be directly monitored through photons emitted from the cavity, which conserve the momentum and energy of polaritons. The state of the condensate can be directly studied by examining the emitted photons leaking through the microcavity mirror. 
It has theoretically been predicted that the negative branch of the excitation spectrum can be observed in exciton-polariton condensates, first in the limit of the equilibrium condition \cite{keeling}, and then in the non-equilibrium condition with pump and decay \cite{szymanska,wouters,tim2}. It has been expected from these theoretical studies that the dispersion relation of the excitation spectrum including the negative branch should be observed through PL measurement.
 Recently, the high density regime of exciton-polariton condensates, which is more than one hundred times the threshold excitation density for condensation, has been investigated and shown that a new high-energy sideband appears \cite{horikiri2}. As the excitation density increased, a high-energy sideband gradually appears. In Ref. \cite{horikiri2}, the observation was studied in detail by comparing with a theory including nonequilibrium features of the electron-hole-photon system \cite{yamaguchi, yamaguchi2}.

  For high density exciton condensates, overlapping excitons at very high densities above the condensation threshold is expected to cause a screening effect of the Coulomb interaction. Similarly, in the case of exciton-polaritons, i.e. an electron-hole-photon (e-h-p) system, it has been believed that the binding of e-h pairs breaks and the system becomes e-h plasma in the high density limit approaching Mott density. Population inversion occurs and photon lasing has been observed after the second nonlinear increase of PL intensity ~\cite{dang,bajoni2,tsotsis}. However, recent experimental data ~\cite{horikiri2, horikiriNJP} have apparently shown different characteristics to photon lasing in the high excitation density regime and implied the existence of attractive e-h pairs in the high density regime. The implication of this difference to the standard photon lasing is theoretically supported by the results that the e-h binding is not only supplied by the Couloumb interaction but also by the cavity coupling to the photons to retain the e-h pair. Therefore, e-h pairs can survive in high density helped by the strong photonic field ~\cite{kamide1,tim,kamide2,yamaguchi,yamaguchi2}.

Here, we focus on the PL close to condensate energy to see whether strong coupling exists far above the condensation threshold by observing dispersion relation including the negative energy branch.
Our sample is comprised of $12$ GaAs quantum wells (QWs) embedded in an AlAs/AlGaAs distributed Bragg reflector microcavity, which we used in our previous studies \cite{chiwei,utsunomiya,george,george2}. A detuning parameter around zero was chosen at a temperature of $8$ K. We used a $4$ ps Ti:Sapphire pulse laser with a $76$ MHz repetition rate to pump the sample at above-band excitation from the reflection dip of the cavity mode at an incidence angle of 50--6$0^\circ$ to study the high excitation density regime.

\begin{figure}
 \begin{center}
 \rotatebox{0}{
 \scalebox{0.17}{
 \includegraphics{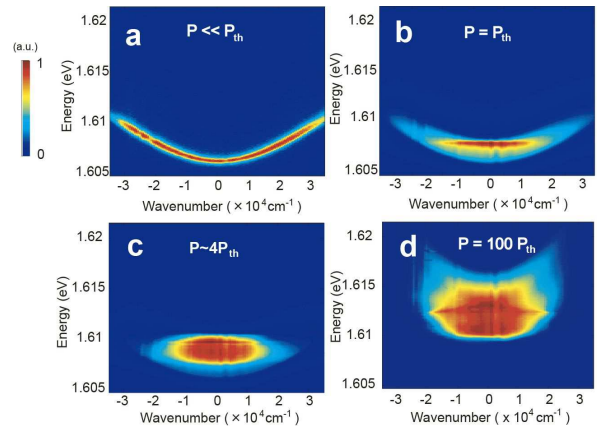}}}
\caption{Time-integrated far-field photoluminescence of exciton-polaritons at several pump powers $P$ as indicated in units of threshold condensation pump power $P_{\rm th}$. All measurements were obtained at 8 K. {\bf a}: Below threshold. {\bf b}: $P_{\rm th}$. {\bf c}: $4P_{\rm th}$. {\bf d}: Far above condensation threshold ($100P_{\rm th}$); negative dispersion branch appears. 
The measurement was done by using a monochromator and a CCD camera, so the repetitive signal of the 76MHz pumped system was accumulated and also the signal inside one period (=1/76MHz~13ns) was summed up. Therefore, high density signal just after the system pumped as well as low density signal (for example, 100ps after pumping) was on the same CCD shot. This is why the linewidth broadening up to 6meV was seen. 
The energy blue shift of the condensate is decreased as the condensate density is reduced. Therefore, the resulting spectrum becomes wider than that for the time-resolved measurements in Fig. \ref{3}.}
\label{2}
 \end{center}
 \end{figure}

Figure \ref{2} shows the measured time-integrated energy-momentum dispersion with increasing excitation density at $\sim$8 K. The dispersion below the condensation threshold (Fig. \ref{2}a) indicates a standard lower polariton (LP) mass that is twice the mass of a bare cavity photon around zero detuning. When the system reaches the condensation threshold, the energy and momentum linewidths decrease as seen in Fig. \ref{2}b, which is associated with a nonlinear increase in PL intensity. The linewidth indicates strong broadening up to $6$ meV at high excitation densities. As the excitation density is continuously increased, the PL emission energy blue shifts, gradually approaching the bare cavity photon energy. Finally, a clear negative branch appears in the spectrum (Fig. \ref{2}d) at an excitation regime of $\sim$$100P_{\rm th}$, where $P_{\rm th}$ denotes the condensation threshold pump rate. The negative branch of dispersion has an effective mass that is estimated to be $m \sim -5 \times 10^{-35}$ kg, which is almost the same value as the LP mass measured below the first threshold. Based on theoretical work including reservoir effects \cite{wouters, tim2}, the excitation spectrum can be transformed from Bogoliubov linear dispersion to quadratic dispersion due to the relaxation oscillation between the condensate and reservoir particles. The excitation density per QW at the pump rate, $P/P_{{\rm th}} \sim 100$, is estimated to be $\sim$$10^{11}\ {\rm cm}^{-2}$. A PL energy approach to the bare cavity photon energy at high densities is expected from the theoretical studies \cite{keeling,kamide1,tim,kamide2,ishida14}. 

Next, we carried out time-resolved spectroscopy of dispersion relations with a streak camera \cite{elena} attached to a monochromator to only examine the time yielding maximum excitation density. 
To avoid reduced time-resolution, the lens in front of the monochromator had a 100 mm focal length introducing a small numerical aperture that caused a small beam spot size at the grating, and hence the temporal dispersion induced by the grating could be reduced to around 10 ps in this study. 


\begin{figure}[htb]
\begin{center}
\scalebox{0.3}{
\includegraphics{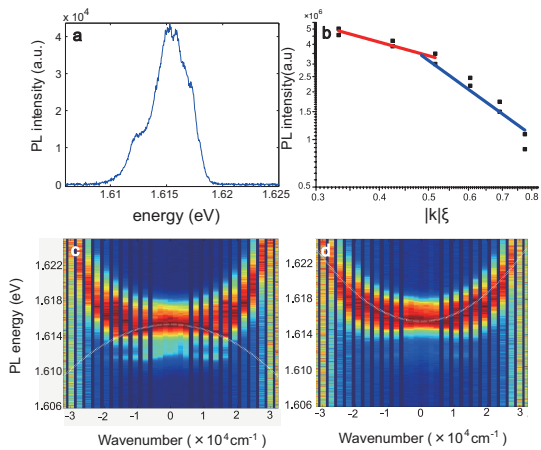}}
\caption{PL energy-momentum relation obtained with time-resolved spectroscopy at far above threshold $P \sim 300 P_{{\rm th}}$. The time for maximum PL intensity was chosen to eliminate the effect of the low density that exists due to the pulsed profile of the pump laser. {\bf a}: Spectrum at $k=0$. {\bf b}: Dependence of PL intensity on $k$. The two points at each $k$ correspond to positive and negative in-plane momenta. The fitting results in the small $k$ (red) regime indicate quantum depletion dominance due to $1/k$ dependence, while the large $k$ regime fitting results (blue) indicate $1/k^2$ dependence. {\bf c} and {\bf d}: Dispersion of the time yielding maximum PL corresponding to the maximum excitation density, which gives a negative branch at around $k=0$. 
 {\bf c}: White dotted curve is the fitting results for the negative branch using a quadratic curve. The curvature provides almost the same absolute value as the normal LP branch below the threshold. PL intensity is normalized at each $k$ value to enable easy comparison with the fitting results. The displayed PL intensities at large $k$ values appear to be as strong as the PL intensity at $k=0$; however, the actual PL is considerably weaker, as seen in Fig. \ref{3}b. {\bf d}: Fitting for positive excitation branch (almost simultaneously with maximum PL but chosen to provide easy fitting to positive branch). The effective mass is that of the normal LP mass below the threshold.}
\label{3}
\end{center}
\end{figure}

Figure \ref{3}a shows a cross section at $k=0$. The main peak around 1.615 eV is the starting point of the negative branch. As $k$ increases, a negative branch is formed. Figure \ref{3}b plots the dependencies of PL intensity on $k$ in the relatively small (red) and large (blue) $k$ regimes in terms of healing length $\xi=\sqrt{\hbar^2\over 2 m_{\rm LP}U(n)}\approx 0.3 \times 10^{-4}\ {\rm cm}$,
where $U(n)$ is the interaction energy (here we use the blue shift energy of 7 meV).
 The slope of the logarithmic plot is approximately $-0.9$ ($\propto$$1/k$) for the small $k$ regime, and this slope value coincides with the population dependence of quantum depletion $n_{\rm quantum}(k)=|v_{-k}|^2= {\hbar^2 k^2/ 2m_{LP}+U(n)\over 2E_{\rm Bog}(k)}- {1\over 2}\rightarrow {\sqrt{m_{\rm LP} U(n)}\over 2\hbar k} \quad \rm{(at \ small }\ \it{k})$, while the slope for large $k$ in the logarithmic plot is approximately $-2.3$ ($\propto 1/k^2$), which is consistent with the effects of thermal depletion, $n_{\rm th}(k)={|u_k|^2+|v_{-k}|^2\over \exp (\beta E_{\rm Bog}(k))-1}\rightarrow {m_{\rm LP}k_B T \over \hbar^2 k^2}\quad \rm{(at \ small }\ \it{k})$. Here, $E_{\rm Bog}(k)=\sqrt{{\hbar^2 k^2\over 2m_{\rm LP}}({\hbar^2 k^2\over 2m_{\rm LP}}+2 U(n)})$ represents the Bogoliubov excitation spectrum and $|u_k|^2 -1/2=|v_{-k}|^2+1/2={\hbar^2 k^2/2m +U(n)\over 2E_{\rm Bog}}$. The value, $|k| \xi \sim 0.5$, corresponds to about $ |k| \sim 1.5 \times 10^4 /{\rm cm}$, which is the regime where the negative branch is bright. This can be fully understood from a scattering model where the negative branch should be bright due to two particles' scattering (quantum depletion) \cite{pieczarka,savvidis} at high densities; however, only the positive branch is populated in the large $k$ regime, possibly due to thermal depletion from the condensate .

The excitation spectrum near $k=0$ can be affected by the presence of reservoir polaritons in the bottleneck region \cite{wouters,tim2}. The effect of the reservoir is to round the linear Bogoliubov dispersion to quadratic dispersion near $k=0$. The dispersion obtained at the time of maximum PL intensity is shown in Fig. \ref{3}c and \ref{3}d. The dispersion mainly indicates a negative branch. The fitting results in Fig. \ref{3}c show a quadratic curve that gives an absolute value close to the effective mass of a normal LP branch below the condensation threshold, while Fig. \ref{3}d shows the fitting results for a positive branch that gives the same effective mass as the LP branch below the threshold (see Supplemental Material for LP dispersion below the threshold). Therefore, the results support the scattering of two particles, which satisfies the conservation of energy and momentum. 
While both positive and negative branches are expected to be observed from the consideration of thermal and quantum depletions, our measured data cannot clearly distinguish two branches due to the broad spectrum. This was not caused by limitations with the resolution of the monochromator attached to the streak camera, which is smaller than 0.1 meV, but possibly by the reduced time-resolution of the streak camera.

\begin{figure}[htb]
\begin{center}
\scalebox{0.45}{
\includegraphics{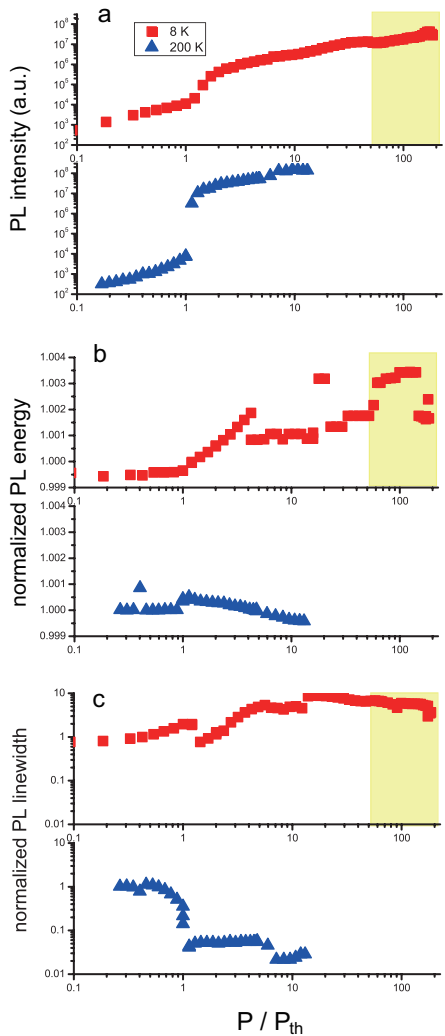}}
\caption{Dependence of PL on excitation for low ($8$ K) and high temperatures ($200$ K). All horizontal axesf values represent pump laser power normalized by condensation and lasing thresholdsf pump power. Negative dispersion can be observed in the region marked in the shaded region (yellow). {\bf a}: PL intensity at $k=0$. The nonlinear increase at the first threshold at $8$ K ($200$ K) indicates the polariton condensation (lasing) threshold. {\bf b}: Maximum PL energy at $k=0$. The blue shift at low temperature is in contrast to the red shift for lasing at $200$ K. {\bf c}: PL linewidth normalized at a linewidth far below the threshold at $k=0$. The sudden narrowing of the linewidth at $P_{\rm th}$ is a typical feature of phase transition. While a laser stays in a narrow linewidth above the threshold, the low temperature case indicates a broader spectrum due to particle-particle interactions.}
\label{4}
\end{center}
\end{figure}

A polariton condensate normally features a second threshold in the high density regime \cite{dang,bajoni,nelsen,bajoni2}. However, the second threshold is not direct evidence of the transition to photon lasing as shown in recent works \cite{yamaguchi2}. 
Our experimental results show a distinct difference from the conventional second-threshold behavior, which shows a lack of a clear second non-linear PL intensity increase as shown in Fig. \ref{4}.
Note that the shaded (yellow) background area in the figure is the regime where we observed the negative branch. It is also characterized by increased linewidth in the regime.

Due to the structural similarities of our sample to a vertical-cavity surface-emitting laser (VCSEL), it is possible to make a crossover from the exciton-polariton condensate to the photon laser by increasing the temperature of the sample. A comparison in the same sample can be made by showing conventional lasing at 200 K. The effective mass measured from the dispersion characteristics below the threshold at $200$ K is $m\sim 3\times10^{-35}$ kg (see Supplemental Material), which is in good agreement with the bare cavity photon mass, thereby indicating the absence of strong coupling. At $ 200$ K (triangles in Fig. \ref{4}) above the threshold, the PL energy exhibits a slight red shift in contrast to the blue shift observed for polariton condensation at $8$ K (red squares in Fig. \ref{4}). The red shift above the laser threshold is due to increased temperature and the absence of particle-particle interactions. The narrowing linewidth observed at $200$ K is also in contrast to the increased linewidth for the case at $8$ K (Fig. \ref{4}c). Note that the broad linewidth of 8 K is partly due to pulsed pump and time-integrate detection; however, as presented in Ref. \cite{horikiriNJP}, the linewidth in the high excitation regime at 8 K is still much broader compared with that at 200 K.

In conclusion, we experimentally studied the high excitation density regime of exciton-polariton condensates. A negative dispersion branch of the PL spectrum as well
as a high-energy side band studied in a previous paper \cite{horikiri2} was directly observed by time-resolved spectroscopy.  This is an indication that strong coupling in the e-h-p system is present in the high density regime.  The negative branch observation, which is considered to be evidence of Bogoliubov particles in the system, would not be attained by alternative theories which decouple the matter fields (electrons and holes) from the photons.  The features seen are distinct from those of a standard photon lasers in the high density regime of past works and the high temperature data of the present work.  This opens the possibility to observe a new regime in the high density e-h-p system, where genuine coherence between the photon and matter fields are present, in contrast to a standard laser where there is no coherence. 
The presence of such coherence is a prerequisite for observing polariton BCS physics \cite{keeling, kamide1, tim,yamaguchi2}, where the quantum state can no longer be understood by a simple macroscopic population of exciton-polaritons.  In order to truly probe the e-h states of the polariton system, a detection that distinguishes between the electrons and holes should be devised to probe their relative correlations. As PL is a recombination of the e-h, quantities such as the relative momentum distribution are unobservable.  Measuring the relative momentum distribution would reveal more information about the high density e-h-p regime. 

The authors wish to thank T. Ogawa, M. Kuwata-Gonokami, J. Keeling, I. Carusotto, K. Kamide, M. Yamaguchi, G. Roumpos, and K. Yan for their helpful comments. This research was supported by the Japan Society for the Promotion of Science (JSPS) through its FIRST Program and KAKENHI Grant Numbers 24740277 and 25800181, a Space and Naval Warfare Systems (SPAWAR) Grant Number N66001-09-1-2024, the Ministry of Education, Culture, Sports, Science, and Technology (MEXT), the National Institution of Information and Communication Technology (NICT), the joint study program at Institute for Molecular Science. T. H. acknowledges the support of  Toray Science Foundation, KDDI Foundation, the Asahi Glass Foundation, ther Murata Science Foundation, and REFEC. T.B. acknowledges the support of the Shanghai Research Challenge Fund, New York University Global Seed Grants for Collaborative Research, National Natural Science Foundation of China (Grant No. 61571301), the Thousand Talents Program for Distinguished Young Scholars (Grant No. D1210036A), and the NSFC Research Fund for International Young Scientists (Grant No. 11650110425).

\end{document}